\documentclass[a4paper,twocolumn,pra,showpacs,showkeywords,aps,nofootinbib]{revtex4}
\newcommand{\beq}{\begin{equation}}
\newcommand{\eeq}{\end{equation}}
\newcommand{\beqa}{\begin{eqnarray}}
\newcommand{\eeqa}{\end{eqnarray}}

\def\ra{\rangle}
\def\la{\langle}

\usepackage{amsmath,amsfonts,amssymb}
\usepackage{graphicx}
\usepackage{bm}

\begin{document}
\title{Shortcuts to adiabaticity for non-Hermitian systems}

\author{S. Ib\'a\~{n}ez$^{1}$}
\author{S. Mart\'\i nez-Garaot$^{1}$}
\author{Xi Chen$^{1,2}$}
\author{E. Torrontegui$^{1}$}
\author{J. G. Muga$^{1}$}

\affiliation{$^{1}$Departamento de Qu\'{\i}mica F\'{\i}sica, Universidad del Pa\'{\i}s Vasco - Euskal Herriko Unibertsitatea,
Apdo. 644, Bilbao, Spain}

\affiliation{$^{2}$Department of Physics, Shanghai University, 200444 Shanghai, China}
\begin{abstract}
Adiabatic processes driven by non-Hermitian, time-dependent Hamiltonians
may be sped up by generalizing inverse engineering techniques based on Berry's transitionless driving algorithm or on dynamical invariants.
We work out the basic theory and examples described by two-level Hamiltonians:
the acceleration of rapid adiabatic passage with a decaying excited level and of the dynamics of a
classical particle on an expanding harmonic oscillator.
\end{abstract}
\maketitle
\section{Introduction}
We refer to fast time-dependent
processes that reproduce the effect of a slow, adiabatic
driving of a quantum system as ``shortcuts to adiabaticity'' \cite{ion,David,MN08,Muga09,S09a,S09b,Berry09,Salamon09,Calarco09,Ch10,ChenPRL10b,ChenET10,Nice10,MN10,Muga10,Li10,Nice11,Chen11,optimal_control,Wu11,Adol11,transport,transport2}.
We also apply the term to the inverse engineering methods used to design these processes. In the adiabatic process of reference the external control parameters are modified slowly
from some initial configuration to a final one. 
In the corresponding shortcut the system is driven in a predetermined short
time to a final state which reproduces in the instantaneous basis the initial populations,
as the adiabatic process would do, but possibly allowing for some transient excitation along the way. There is nowadays considerable interest in these questions for
fundamental and practical reasons. Adiabatic methods are ubiquitous
in cold-atom and atomic-physics laboratories
to manipulate and prepare atomic states in principle in a robust way.
An obvious drawback is that the times required may be too long for practical applications. Moreover the ideal robustness may be spoiled by the accumulation of perturbations and decoherence due to noise and undesired interactions.
Studies and experiments to speed up adiabatic processes have been carried out for transport \cite{ion,David,MN08,Calarco09,transport,transport2}, wave splitting \cite{S09a,S09b}, expansions
and compressions \cite{Muga09,Salamon09,Ch10,Muga10,ChenET10,MN10,Nice10,Nice11,Li10,optimal_control,Wu11,Adol11}, 
or internal state control
\cite{Berry09,ChenPRL10b,Chen11}.
These studies have so far been performed for Hermitian Hamiltonians, but many systems admit an effective non-Hermitian description.
In this paper we put forward shortcuts to adiabaticity techniques for non-Hermitian Hamiltonians. Specifically we shall generalize the
inverse engineering method proposed by Berry \cite{Berry09}
and the one based on dynamical invariants \cite{Ch10}.
While these methods are intimately connected as shown in \cite{Chen11} and may in fact be
considered potentially equivalent, in standard applications they are used
in different ways and provide different answers so we shall consider them separately here.
As study cases we shall discuss
a two-level decaying atom and the motion of a classical particle in a
harmonic oscillator with time-dependent frequency.
%
%
%
%
\subsection{Non-Hermitian Hamiltonians: basic formulae\label{nhh}}
Non-Hermitian Hamiltonians typically describe subsystems of a larger system
\cite{Muga04}.
We shall first review a basic set of relations and notation \cite{Muga04}.
We shall assume a non-Hermitian time-dependent Hamiltonian $H_0(t)$ with $N$ non-degenerate right eigenstates $\{|n(t)\ra\}$, $n=1,2...,N$,
\beq
\label{eigenvalues_eq}
H_0(t) |n(t)\ra = E_n(t) |n(t)\ra,
\eeq
and biorthogonal partners $\{|\hat{n}(t)\ra\}$,
\beq
\label{eigenvalues_eq1}
H^{\dag}_0(t)  |\hat{n}(t)\ra = E_n^{*}(t)  |\hat{n}(t)\ra, 
\eeq
where the star means ``complex conjugate'' and the dagger denotes the adjoint
operator.   
They satisfy
\beq
\label{delta}
\la\hat{n}(t)|m(t)\ra = \delta_{nm}
\eeq
and the closure relations
\beq
\sum_n |\hat{n}(t)\ra \la n(t)| = \sum_n |n(t)\ra\la \hat{n}(t)|=1.
\eeq
%
$\la\hat{n}(t)|$ is the left eigenvector of $H_0(t)$,
\beq
\la\hat{n}(t)| H_0(t)= \la\hat{n}(t)| E_n(t),
\eeq
and $\la n(t)|$ the left eigenvector of $H^{\dag}_0(t)$,
\beq
\la n(t)| H^{\dag}_0(t)= \la n(t)|  E_n^{*}(t).
\eeq
We can thus write the Hamiltonian and its adjoint as
\beqa
\nonumber
H_0(t)&=& \sum_n |n(t)\ra E_n(t) \la\hat{n}(t)|,
%
%
\\
H^{\dag}_0(t) &=& \sum_n |\hat{n}(t)\ra E_n^{*}(t) \la n(t)|.
\eeqa
The time-dependent Schr\"{o}dinger equations for
a generic state $|\Psi(t)\ra$ and for its biorthogonal partner
$|\hat{\Psi}(t)\ra$
satisfying $\la \hat{\Psi}(t)|\Psi(t)\ra=1$ are
%
\beqa
\label{Schrodinger_eq}
i\hbar \partial_t |\Psi(t)\ra&=& H_0(t) |\Psi(t)\ra,
\\
%
%
\label{adjoint_eq}
i\hbar \partial_t |\hat{\Psi}(t)\ra&=& H^{\dag}_0(t) |\hat{\Psi}(t)\ra.
\eeqa
%
%
%
%
%
%
%
%
%
%
%
%
%
\section{Transitionless driving algorithm}
%
%
%
%
%
%
%
In \cite{Berry09} M. V. Berry proposed a method to design a Hermitian Hamiltonian $H(t)$ for which the approximate adiabatic dynamics driven by the
Hermitian Hamiltonian $H_0(t)$ becomes exact. We shall generalize this method for non-Hermitian
Hamiltonians. First we need the adiabatic approximation
when $H_0(t)$ is non-Hermitian \cite{adNH1,adNH2}.
A general time-dependent state $|\Psi(t)\ra$
is a linear combination of instantaneous eigenvectors $|n(t)\ra$
of $H_0(t)$ with time-dependent coefficients. Similarly $|\hat\Psi(t)\ra$
is a linear combination of instantaneous eigenvectors $|\hat{n}(t)\ra$
of $H^\dagger_0(t)$.
In the adiabatic approximation we assume that only one of these eigenvectors is populated.
To determine the corresponding phase factor
%
%
%
%
%
we insert
\beqa
\label{psi_n}
|\varphi_n(t)\ra &=& e^{i \beta_n(t)} |n(t)\ra,
%
%
\\
\label{psi_n1}
|\hat{\varphi}_n(t)\ra &=& e^{i \hat{\beta}_n(t)} |\hat{n}(t)\ra,
\eeqa
%
into Eqs. (\ref{Schrodinger_eq}) and
(\ref{adjoint_eq}). Thus we have
\beqa
\label{eq}
i \dot{\beta}_n |n(t)\ra + |\partial_t n(t)\ra= \frac{H_0(t)}{i\hbar}  |n(t)\ra,
%
%
\\
\label{eq1}
i \dot{\hat{\beta}}_n |\hat{n}(t)\ra + |\partial_t \hat{n}(t)\ra= \frac{H^{\dag}_0(t)}{i\hbar}  |\hat{n}(t)\ra,
\eeqa
where the dot denotes the derivative with respect to time. Multiplying Eq. (\ref{eq}) by $\la\hat{n}(t)|$ and Eq. (\ref{eq1}) by $\la n(t)|$,
taking into account Eqs. (\ref{eigenvalues_eq}) and (\ref{eigenvalues_eq1}),
and integrating, we find
%
%
%
%
%
%
\beqa
\label{phase}
\beta_n(t)&=& \int^t_0 \left[ \frac{-E_n(t')}{\hbar}
+i \la\hat{n}(t')|\partial_{t'} n(t')\ra\right] dt',
\\
%
%
\label{phase1}
\hat{\beta}_n(t)&=& \int^t_0 \left[  \frac{-E^*_n(t')}{\hbar}+i \la n(t')|\partial_{t'} \hat{n}(t')\ra\right]dt',
\eeqa
where the initial phases are set to zero.
As $\la n(t)|\partial_{t}\hat{n}\ra^* = \la\partial_{t}\hat{n}|n(t)\ra$ and, from Eq. (\ref{delta}),
$\la\partial_{t}\hat{n}|n(t)\ra = -\la\hat{n}(t)|\partial_{t}n\ra$, we have
that $\hat{\beta}_n=\beta_n^*$.

As in \cite{Berry09},
we now impose that
all $|\varphi_n(t)\ra$ satisfy exactly the Schr\"{o}dinger equation for a yet unknown $H(t)$,
\beq
i\hbar \partial_t |\varphi_n(t)\ra= H(t) |\varphi_n(t)\ra.
\eeq
Similarly,
\beq
i\hbar \partial_t |\hat{\varphi}_n(t)\ra= H^{\dag}(t) |\hat{\varphi}_n(t)\ra.
\eeq
%
%
The states $|\varphi_n(t)\ra$ and $|\hat{\varphi}_n(t)\ra$ can be written in terms of the corresponding evolution operators $U(t)$ and
$\hat{U}(t)$,
\beqa
\nonumber
|\varphi_n(t)\ra&=& U(t) |n(0)\ra,
%
%
\\
\label{evolution1}
|\hat{\varphi}_n(t)\ra&=& \hat{U}(t) |\hat{n}(0)\ra.
\eeqa
The Hamiltonian $H(t)$ can be found from
\beq
i\hbar \partial_t U(t)= H(t) U(t),
\eeq
as
\beq
\label{H}
H(t)= i\hbar \partial_t U(t) \hat{U}^{\dag}(t),
\eeq
since \cite{Muga04}
\beq
\hat{U}^{\dag}(t) U(t)= 1_{op}.
\eeq
%
The evolution operators can
be written as
\beqa
U(t) &=& \sum_n e^{i \beta_n(t)} |n(t)\ra \la\hat{n}(0)|,
\nonumber
%
%
\\
\hat{U}(t)&=& \sum_n e^{i \hat{\beta}_n(t)} |\hat{n}(t)\ra \la n(0)|.
\eeqa
Using now Eq. (\ref{H}),
\beq
\label{Berry}
H(t) = H_0(t)+H_1(t),
\eeq
%
%
where
\beqa
\label{H1}
H_1(t) &=& i\hbar   \sum_n [ |\partial_{t} n(t)\ra\la\hat{n}(t)|
\nonumber \\
&-& \la\hat{n}(t)|\partial_{t} n(t)\ra   |n(t)\ra\la\hat{n}(t)|].
\eeqa
$H(t)$ drives the system
along the adiabatic paths
defined by $H_0(t)$.

As noted in \cite{Berry09} and \cite{Chen11} this Hamiltonian is not unique.
For a given set $|n(t)\ra$ the same final populations are
found by choosing different phases.
Let us rewrite
$|\varphi_n(t)\ra$ and
$|\hat{\varphi}_n(t)\ra$ in terms of arbitrary phases, $\xi_n(t)$ and $\hat{\xi}_n(t)$,  which we now consider manipulable functions obeying $\xi_n(t)= \hat{\xi}_n^*(t)$ so that $\la\hat{\varphi}_n(t)|\varphi_n(t)\ra=1$,
%
\beq
|\varphi_n(t)\ra= e^{i\xi_n(t)} |n(t)\ra,\,\,
|\hat{\varphi}_n(t)\ra= e^{i\hat{\xi}_n(t)} |\hat{n}(t)\ra.
\eeq
%
%
%
We assume
$\xi_n(0)= \hat{\xi}_n(0)= 0$ and define the new evolution operators
\beqa
U_{\xi}(t) &=& \sum_n e^{i \xi_n(t)} |n(t)\ra \la\hat{n}(0)|,
\nonumber\\
%
%
\hat{U}_{\xi}(t)&=& \sum_n e^{i \hat{\xi}_n(t)} |\hat{n}(t)\ra \la n(0)|.
\eeqa
From Eq. (\ref{H}), the corresponding Hamiltonian becomes
\beq
\label{Hxi}
H_{\xi}(t)= -\hbar \sum_n |n(t)\ra \dot{\xi}_n(t) \la\hat{n}(t)| + i\hbar   \sum_n |\partial_{t}n(t)\ra\la\hat{n}(t)|.
\eeq
%
%
%
%
%
%
%
%
%
%
%
\section{Transitionless driving algorithm applied to a decaying
two-level atom}
\subsection{$H_{1}(t)$ applied to a decaying two-level atom}
As an example of the approach of the previous section we shall speed up
adiabatic processes in a two-level atom
with spontaneous decay. If the decayed atom escapes from the trap by recoil, 
a Hamiltonian (rather than master equation) description is enough.
We shall also assume a semiclassical treatment of the interaction between a 
laser electric field
linearly polarized in $x$-direction,
and a  decay rate (inverse life-time) $\Gamma$ from the excited state.

Applying the electric dipole approximation, a laser-adapted interaction picture and 
the rotating wave approximation, the Hamiltonian 
is 
\beq
H_{a0}(t)=\frac{\hbar}{2}
\left(\begin{array}{cc}
-\Delta(t) & \Omega_{R}(t)\\
\Omega_{R}(t) & \Delta(t)-i\Gamma
\end{array} \right),
\eeq
in the atomic basis $|1\rangle = \left( \begin{array} {rccl} 1\\ 0 \end{array} \right)$, $|2\rangle = \left( \begin{array} {rccl} 0\\ 1 \end{array} \right)$.
The detuning from the atomic transition frequency $\omega_0$ is
$\Delta(t)=\omega_0-\omega_i(t)$, where $\omega_i(t)$ is the instantaneous field frequency.
We assume a slowly varying pulse envelope so that 
the Rabi frequency
$\Omega_R(t)$, assumed real, depends on time. 
In the example below we shall take $\Gamma$ as a constant although,
in a general case, it could also depend on time, $\Gamma=\Gamma(t)$, as an effective decay rate controlled by further interactions, see e.g. \cite{Zeno}.    
The eigenvalues of this Hamiltonian are
\beq
\label{E_+_-}
E_{\pm}(t) = \frac{\hbar}{4} \left\{-i\Gamma \pm \sqrt{-[\Gamma+2i\Delta(t)]^2 + 4\Omega^2_R(t)} \right\},
\eeq
and the normalized eigenstates are
\beqa
\nonumber
|\chi_{+}(t)\ra = \sin\left(\frac{\alpha}{2}\right) |1\ra + \cos\left(\frac{\alpha}{2}\right) |2\ra,
\\
\label{n-}
|\chi_{-}(t)\ra = \cos\left(\frac{\alpha}{2}\right) |1\ra - \sin\left(\frac{\alpha}{2}\right) |2\ra,
\eeqa
%
%
%
%
%
where the mixing angle $\alpha=\alpha(t)$ is complex and defined as
\beq
\tan \alpha = \frac{\Omega_R}{\Delta - i \Gamma}.
\eeq
%
%
%
%
%
%
%
%
%
The adjoint of $H_{a0}(t)$ is
\beq
H_{a0}^{\dag}(t)=\frac{\hbar}{2}
\left(\begin{array}{cc}
-\Delta(t) & \Omega_{R}(t)\\
\Omega_{R}(t) & \Delta(t)+i\Gamma
\end{array} \right),
\eeq
with eigenvalues $E_{\pm}^*(t)$ and normalized eigenstates
\beqa
\nonumber
|\hat\chi_{+}(t)\ra = \sin\left(\frac{\alpha^*}{2}\right) |1\ra + \cos\left(\frac{\alpha^*}{2}\right) |2\ra,
\\
\label{n-1}
|\hat\chi_{-}(t)\ra= \cos\left(\frac{\alpha^*}{2}\right) |1\ra - \sin\left(\frac{\alpha^*}{2}\right) |2\ra.
\eeqa
%
%
%
Note that the coefficients are complex conjugate of those in 
Eq. (\ref{n-}) because 
$H_{a0}(t)$ is equal to its transpose \cite{Muga04}.
For this system Eq. (\ref{H1}) takes the form
\beqa
H_{a1}(t) &=& i\hbar [|\partial_{t} \chi_{+}(t)\ra\la\hat{\chi}_{+}(t)|
\nonumber \\
&-& \la\hat{\chi}_{+}(t)|\partial_{t} \chi_{+}(t)\ra |\chi_{+}(t)\ra\la\hat{\chi}_{+}(t)|
\nonumber \\
&+& |\partial_{t} \chi_{-}(t)\ra\la\hat\chi_{-}(t)|
\nonumber \\
&-& \la\hat{\chi}_{-}(t)|\partial_{t} \chi_{-}(t)\ra |\chi_{-}(t)\ra\la\hat\chi_{-}(t)|],
\eeqa
where, according to Eqs. (\ref{n-}) and (\ref{n-1}),
\beqa
\nonumber
\la\hat\chi_{\pm}(t)|\partial_{t} \chi_{\pm}(t)\ra &=& 0,
\\
\la\hat\chi_{\mp}(t)|\partial_{t} \chi_{\pm}(t)\ra &=& \pm \frac{\dot{\alpha}}{2} ,
\eeqa
so
\beq
H_{a1}(t)= \hbar
\left(\begin{array}{cc}
0  &  C(t)  \\
-C(t)   &   0
\end{array} \right),
\label{Ha1}
\eeq
where $C(t)= i \dot{\alpha}/2$ and
\beqa
\dot{\alpha} = \frac{\dot{\Omega}_R [\Delta(t) - i \Gamma/2]- \Omega_R(t) (\dot{\Delta} - i \dot{\Gamma}/2)}{[\Delta(t)-i \Gamma/2]^2 + \Omega^2_R(t)}.
\eeqa
Then,
the Hamiltonian $H_a(t)=H_{a0}+H_{a1}$ takes the form
\beq
H_a(t)= \frac{\hbar}{2}
\left(\begin{array}{cc}
-\Delta(t)  &  \Omega_R(t)+2C(t)  \\
\Omega_R(t)-2C(t)   &   \Delta(t)-i\Gamma
\end{array} \right).
\eeq
The practical realization of this Hamiltonian is not straightforward.
In particular the off-diagonal terms are not the complex conjugate of each other
unless the real part of $C(t)$ becomes zero.
We shall explore in the following subsection the possibility to manipulate this result by playing
with different phases as in Eq. (\ref{Hxi}). 
%
%
%
%
%
%
%
%
%
%
%
%
\subsection{$H_{\xi}(t)$ applied to a decaying two-level atom}
{}For the decaying two-level atom, using Eq. (\ref{Hxi})
with phases $\xi_{+}= \xi_{+}(t)$ and $\xi_{-}= \xi_{-}(t)$
associated with $|\chi_{+}(t)\ra$ and $|\chi_{-}(t)\ra$,
we find
%
%
%
%
\beqa
\nonumber
&& \frac{H_{\xi a}(t)}{\hbar}=
\\ \nonumber
&&\!\!\!\!\!\!\left[\!\!\begin{array}{cc}
-\sin^2\!{\left(\frac{\alpha}{2}\right)} \dot{\xi}_{+}- \cos^2\!{\left(\frac{\alpha}{2}\right)} \dot{\xi}_{-}\!  &\!
\frac{\sin{\alpha}}{2} (\dot{\xi}_{-} - \dot{\xi}_{+}) + C   \\
\frac{\sin{\alpha}}{2} (\dot{\xi}_{-} - \dot{\xi}_{+})- C\!   &\!   -\cos^2\!{\left(\frac{\alpha}{2}\right)} \dot{\xi}_{+}-\sin^2\!{\left(\frac{\alpha}{2}\right)} \dot{\xi}_{-}
\end{array}\!\!\right]\!\!.
\\
\eeqa
%
The phases in the matrix elements $H_{\xi a,12}(t)$ and $H_{\xi a,21}(t)$ only affect the first terms, which are equal. In general the manipulation of the
phases is not enough to make the non-diagonal terms complex conjugate
of each other since this requires
not only ${\rm{Im}}[(\dot{\xi}_{-} - \dot{\xi}_{+})\sin\alpha]=0$ but
${\rm{Re}}[C(t)]=0$ too. We also add potentially complex terms in the diagonal
that again could complicate the physical realization.
%

In summary, the phase manipulation does not help to implement the shortcut.  
In some parameter regimes, however, an approximation to $H_{a1}$ that leads to
essentially the same results may be easily realized, as discussed next.   
\subsection{Forced population inversion}
We study now the forced coherent decay from the upper level
of a two-level system with slow spontaneous decay.
This decay may be driven and accelerated adiabatically
with a ``rapid'' adiabatic passage (RAP) technique, sweeping the laser frequency across
resonance. The adjective ``rapid'' here could be misleading: it simply means ``faster
than the spontaneous decay'' but, as the approach is adiabatic,
it fails for short enough times. The adiabaticity criterion is worked out in the appendix.
To go beyond the time limits imposed by the
breakdown of adiabaticity, shortcut techniques may be applied.

%

%
%
%
\begin{figure}[h]
\begin{center}
\includegraphics[height=4.5cm,angle=0]{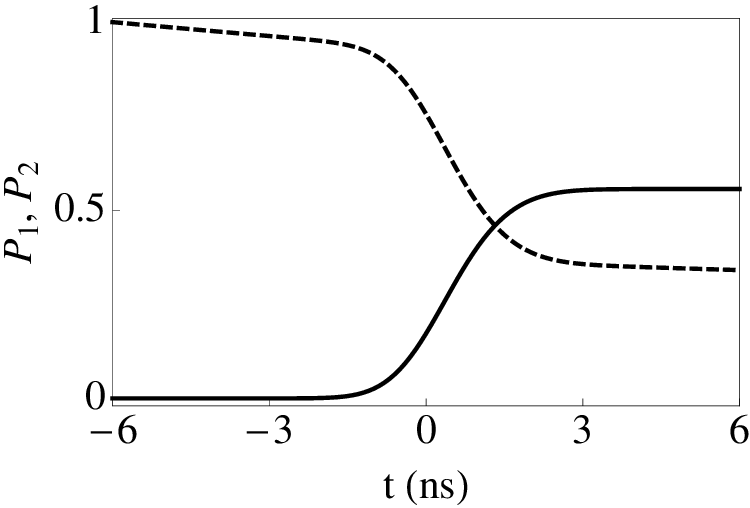}
\end{center}
\caption{\label{P1_P2_H0}
Population of the ground state, $P_1(t)$ (solid line), and of the excited state, $P_2(t)$ (dashed line),
for the Hamiltonian $H_{a0}(t)$.
Parameters: $\Gamma= 2 \pi \times 2$ MHz, $a= (2\pi)^2 \times 0.01$ GHz$^2$, $b= (2 \pi)^2 \times 0.00025$  GHz$^2$, and
$\Omega_0= 2 \pi \times 100$ MHz.}
\end{figure}
%
%

We consider a linearly chirped Gaussian pulse
with detuning
%
%
$\Delta(t)= \omega_0-\omega(t)= -2bt$
and Gaussian Rabi frequency $\Omega_R(t)= \Omega_0 e^{-at^2}$.

\begin{figure}[h]
\begin{center}
\includegraphics[height=4.5cm,angle=0]{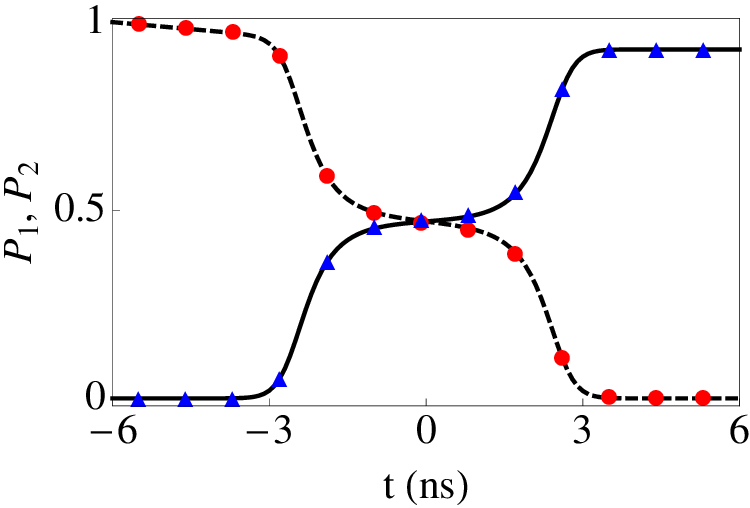}
\end{center}
\caption{\label{P1_P2_H}
(Color online) Population of the ground state, $P_1(t)$ (solid line), and of the excited state,
$P_2(t)$ (dashed line), for the total Hamiltonian $H_{a}(t)$, coinciding with the
populations $P_1(t)$ (triangles) and
$P_2(t)$ (circles) when $H_{a}(t)$ is approximated by neglecting ${\rm{Re}}[C(t)]$.
Parameters as in Fig. \ref{P1_P2_H0}.}
\end{figure}

\begin{figure}[h]
\begin{center}
\includegraphics[height=4.5cm,angle=0]{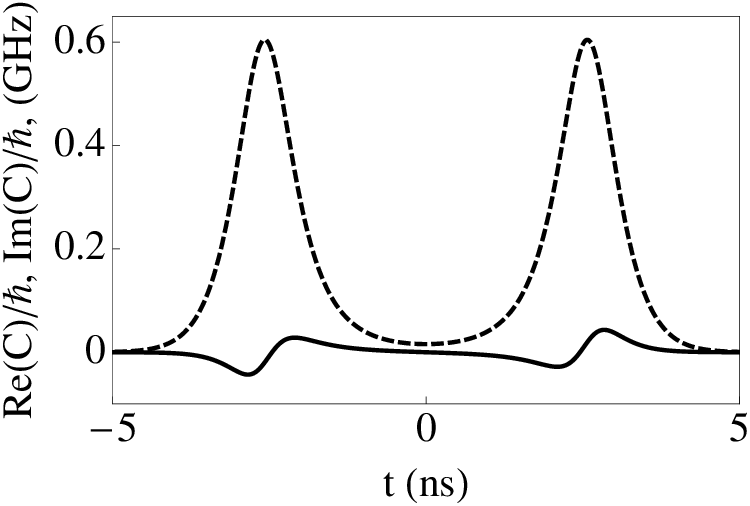}
\end{center}
\caption{\label{reimC}
Real (solid line) and imaginary (dashed line) parts of $C$.
Parameters as in Fig. \ref{P1_P2_H0}.}
\end{figure}

%
The initial conditions are $P_1(0)=0$ and $P_2(0)=1$.
In Fig. \ref{P1_P2_H0} we show that the application of a RAP
pulse with $H_{a0}(t)$ is only partially successful. Note the slow spontaneous
decay before and after the pulse, and a faster forced transition during the pulse
around $t=0$. Since the pulse duration is too short, adiabaticity fails.
Fig. \ref{P1_P2_H} shows the fast full population inversion
when adding the Hamiltonian $H_{a1}(t)$ in Eq. (\ref{Ha1}). This Hamiltonian
has off-diagonal terms with real and imaginary parts depicted in
Fig. \ref{reimC}. Whereas the imaginary parts, the bigger  bumps in
Fig. \ref{reimC}, are realizable 
\cite{ChenPRL10b},
the real parts constitute a non-Hermitian contribution.
They are however small, and an approximation of $H_{a1}(t)$ neglecting
them provides essentially the same dynamics, as shown in
Fig. \ref{P1_P2_H}. This remains valid in the strong-driving regime
in which $\Gamma \ll \Omega_0$ and the natural lifetime is large
compared to the duration of the forced decay.

%
%
%
%
%
%
%
%
%
%
%
%
%
%
%
%
%
\section{Invariants based inverse engineering}
Lewis and Riesenfeld \cite{LR} proposed the use of dynamical invariants of a quantum mechanical system to perform expansions
of arbitrary time-dependent wave functions by superposition of eigenstates of the invariant.
This may be generalized to non-Hermitian Hamiltonians \cite{Gao91,Gao92,Gao93}.
We shall assume that for a Hamiltonian $H_0(t)$ with the features described in Sec. \ref{nhh}, there
is a generalized invariant $I(t)$ that satisfies
\beqa
\frac{\partial{I}(t)}{\partial{t}} - {\frac{i}{\hbar}} [I(t),H_0(t)] = 0,
\label{condinv}
\eeqa
so that $\frac{d}{dt}\la \hat{\Psi}(t)|I(t)|\Psi(t)\ra=0$. Note that this is not an ordinary expectation value $\la\Psi(t)|I(t)|\Psi(t)\ra/\la \Psi(t)|\Psi(t)\ra$, in this sense the concept of generalized invariant differs from the one for Hermitian Hamiltonians.

Let us assume also that $I(t)$ has a non-degenerate complete biorthonormal set of
instantaneous eigenstates, $\{|\psi_{n}(t)\ra,|\hat{\psi}_{n}(t)\ra\}$, where $n$ varies from $1$ to $N$, that satisfy
\beqa
\label{Ieigenvalues_eq}
I(t) |\psi_{n}(t)\ra &=& I_{n}(t) |\psi_{n}(t)\ra,
\\
%
%
\label{Ieigenvalues_eq1}
I^{\dag}(t)  |\hat{\psi}_{n}(t)\ra &=& I_{n}^{*}(t)  |\hat{\psi}_{n}(t)\ra,
\\
%
%
\la\hat{\psi}_{m}(t)|\psi_{n}(t)\ra &=& \delta_{m n},
\\
%
%
\sum_{n} |\hat{\psi}_{n}(t)\ra \la\psi_{n}(t)| &=& 1.
\eeqa
We can write the general solutions of the Schr\"{o}dinger equations for $H_0(t)$ and $H_0^{\dag}(t)$, Eqs. (\ref{Schrodinger_eq}) and
(\ref{adjoint_eq}), as
\beqa
|\Psi(t)\ra&=& \sum_n d_n e^{i \alpha_{n}(t)} |\psi_{n}(t)\ra,
%
%
\\
|\hat{\Psi}(t)\ra&=&\sum_n \hat{d}_n e^{i \alpha_{n}^{*}(t)} |\hat{\psi}_{n}(t)\ra,
\eeqa
where the coefficients $\{d_n\}$ and $\{\hat{d}_n\}$ do not depend on time,
and the generalized Lewis-Riesenfeld phases are
\beq
\alpha_{n}(t)=  \int_0^t \bigg\la\hat{\psi}_{n}(t')\bigg| i \frac{\partial}{\partial{t'}}-H(t') \bigg|\psi_{n}(t')\bigg\ra dt'.
\eeq
Inverse engineering techniques rely on designing the invariant eigenvectors
and phase factors first, possibly taking into account partial
information on the structure of the Hamiltonian,
and then deducing the Hamiltonian from them.
\section{Classical particle in an expanding harmonic trap}
It is possible to study a classical particle 
with position $q(t)$ and momentum $p(t)$ in a harmonic trap as a formal quantum two-level system with non-Hermitian Hamiltonian, by rewriting the classical canonical equations of motion in matrix form \cite{Gao91,Gao92}. 
%
%
The Hamiltonian of a classical harmonic oscillator with a time dependent frequency $\omega(t)$ is
\beq
\label{H_ho}
H_{ho}(t)=  \frac{p^2}{2m} + \frac{1}{2} m\omega^2(t) q^2,
\eeq
where $m$ is the mass of the particle. We shall consider an expansion from $\omega_0= \omega(0)$ at $t=0$ to $\omega_f= \omega(t_f)$ at the final time $t=t_f$, 
with $\omega_f < \omega_0$.
The corresponding classical canonical equations
\beqa
\dot{q}_j &=& \frac{\partial{H_{ho}}}{\partial{p}_j} = \frac{p(t)}{m}, 
%
%
\\
\dot{p}_j &=& -\frac{\partial{H_{ho}}}{\partial{q}_j} =-m \omega^2(t) q(t),
\eeqa
can be written as
\beq
\left(\begin{array}{cc} 
\dot{q}  \\
\dot{p}
\end{array} \right)   =
\left(\begin{array}{cc} 
0   &   1/m      \\
-m \omega^2(t)   &   0
\end{array} \right)
\left(\begin{array}{cc} 
q(t)  \\
p(t)
\end{array} \right),
\eeq
due to their linear dependence on $q$ and $p$. Multiplying both sides of the equality by $i$ we obtain a Schr\"odinger-like equation ($\hbar=1$) with the ``effective'' non-Hermitian
Hamiltonian
\beq
\label{Ham}
\mathcal{H}(t)= i
\left(\begin{array}{cc} 
0   &   1/m      \\
-m \omega^2(t)   &   0
\end{array} \right).
\eeq
%
This is a useful but formal analogy, since this Hamiltonian does not
have units of energy, in fact different matrix elements have different units
as the ``state vector'' components $q$ and $p$ have also different units.   
Nevertheless we may apply the generalized invariant theory, and expand the state vector
in terms of formal eigenvectors of the generalized invariants. 
Defining  
\cite{Gao92}
\beq
\label{I}
\mathcal{I}(t)= 
\left(\begin{array}{cc} 
b(t)   &   c(t)      \\
-a(t)   &   -b(t)
\end{array} \right),
\eeq
%
and imposing Eq. (\ref{condinv}), without $\hbar$,   
\beqa
\label{a}
a(t)&=&m  \left[ \frac{\omega_0}{\varrho^2(t)} + \frac{1}{\omega_0} \dot{\varrho}^2(t) \right],
\\
\label{b}
b(t)&=&\frac{-1}{\omega_0 }\varrho(t) \dot{\varrho}(t),
\\
%
\label{c}
c(t)&=&\frac{\varrho^2(t)}{\omega_0 m},
\eeqa
where the dimensionless scaling function $\varrho(t)$ 
satisfies the auxiliary equation 
\beq
\label{ro}
\ddot{\varrho}(t) + \omega^2(t) \varrho(t) = \frac{\omega^2_0}{\varrho^3(t)},
\eeq
which is the Ermakov equation, the same 
equation for the scaling function that defines the invariants 
in the expansion of the 
quantum harmonic oscillator \cite{Ch10}. 
For $\mathcal{I}(t)$, whose eigenvalues are
$I_{\pm}= \mp i$, the eigenstates are
\beq
|\psi_{\pm}(t)\ra=
\left(\begin{array}{cc} 
c(t)                \\
\pm i \left[ 1 \pm i b(t)\right]   
\end{array} \right),
\eeq
in the basis used in Eq. (\ref{Ham}). The Lewis-Riesenfeld phases $\alpha_{\pm}(t)$ are
\beqa
\alpha_{\pm}(t) &=&  \int_0^t \bigg\la\hat{\psi}_{\pm}(t')\bigg| i \frac{\partial}{\partial{t'}}-\mathcal{H}(t') \bigg|\psi_{\pm}(t')\bigg\ra dt'
\nonumber \\
&=& i \ln{\sqrt{\frac{c(t)}{c(0)}}} \pm  \omega_0 \int_0^t  \frac{1}{\varrho^2(t')}  dt' .
\eeqa
Then, the phase-space trajectory is given by
\beqa
\left(\begin{array}{cc} 
q(t)   \\
p(t)
\end{array} \right)
&=&
d_{+} e^{i \alpha_{+}(t)} |\psi_{+}(t)\ra + d_{-} e^{i \alpha_{-}(t)} |\psi_{-}(t)\ra
\nonumber \\
&=&\!R\!
\left(\!\!\begin{array}{cc} 
\varrho(t) \cos{\theta(t)}    \\
- \frac{m \omega_0}{\varrho(t)} \sin{\theta(t)}  + m \dot{\varrho}(t) \cos{\theta(t)}
\end{array}\!\!\right)\!,
\eeqa
where $R= 2r \sqrt{c(0)/m\omega_0}$ is a distance, $d_{+}= d_{-}^*= r \exp{(i \theta_0)}$ can be determined by the initial conditions
at $t=0$, and
\beqa
\theta(t) = \omega_0 \int_0^t  \frac{1}{\varrho^2(t')}  dt' + \theta_0,
\eeqa
with $\theta_0$ the initial phase.

Imposing the boundary conditions $\varrho(0)=1$ and $\dot{\varrho}(0)=0$,
and $\varrho(t_f)=(\omega_0/\omega_f)^{1/2}$ and $\dot{\varrho}(t_f)=0$, which consistently with the Ermakov equation imply $\ddot{\varrho}(0)=\ddot{\varrho}(t_f)=0$,
we find $E_0:=E(t=0)=\omega_0^2R^2m/2$ and $E_f:=E(t_f)=\omega_f E_0/\omega_0$.
In other words,
these boundary conditions guarantee that the value of the classical adiabatic invariant 
$E(t)/\omega(t)$ at initial and final times coincides, even though it may 
take different values at intermediate times.

To design the process, $\varrho(t)$ has to be interpolated at intermediate times.  
We assume here a polynomial form, $\varrho(t)= \sum_{n=0}^{5} a_n t^n$,
where the coefficients $a_n$ are fixed from the boundary conditions. 
Then we get $\omega(t)$ from the Ermakov equation,   
\beqa
\label{omega_t}
\omega(t)= \sqrt{\frac{\omega_0^2}{\varrho^4(t)}-\frac{\ddot{\varrho}(t)}{\varrho(t)}}.
\eeqa

\begin{figure}[h]
\begin{center}
\includegraphics[height=4.5cm,angle=0]{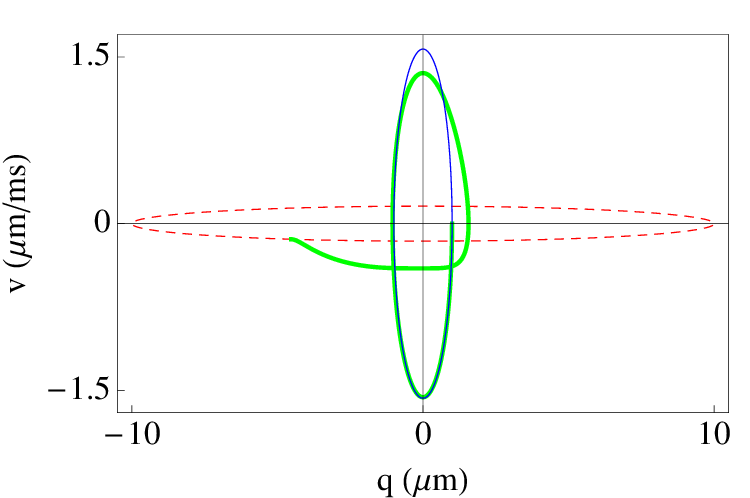}
\end{center}
\caption{\label{traj}
(Color online) Parametric velocity-position trajectory.  
The initial ellipse (solid blue thin line) and the final ellipse (red dashed line) 
are connected by the
shortcut trajectory (green solid thick line). 
Parameters: $\theta_0=0$ corresponding to $q_0=q(t=0)=1$ $\mu$m and $v_0=v(t=0)=0$ $\mu$m/ms, $\omega_0= 2 \pi \times 250$ Hz,
$\omega_f=2 \pi \times 2.5$ Hz, $t_f= 25$ ms, and the mass of an atom of Rubidium-87, $m=1.44 \times 10^{-25}$ kg.}
\end{figure}

In Fig. \ref{traj}
we have represented the shortcut trajectory in phase space between the initial and final times, $t=0$ and $t=t_f$, for the frequency
$\omega(t)$ given by Eq. (\ref{omega_t}). We have also added a period 
$T_0= 2\pi/\omega_0$ before $t=0$, and a period $T_f= 2\pi/\omega_f$ after $t=t_f$, for which  the particle evolves 
for fixed $\omega_0$ and $\omega_f$, respectively, 
so as to depict complete initial and final ellipses.    
The shortcut trajectory that connects the initial and final ellipses 
is clearly not an adiabatic path, that would be formed by a succession of
slowly varying ellipses from the initial to the final one.

%
%

%
%
%
%
%
%
\section{Discussion and Conclusion}
We have generalized shortcut to adiabaticity techniques for non-Hermitian Hamiltonian systems
and provided application examples. Experimental implementations are at reach.
Related open questions are the application of similar
concepts to master equations, or developing means to implement arbitrary
non-Hermitian interactions.
Another interesting research avenue is to combine shortcut techniques with optimal control \cite{Li10,optimal_control} taking into account physically imposed constraints.
\section*{Acknowledgments}
We acknowledge
funding by the Basque Government (Grant No. IT472-10)
and Ministerio de
Ciencia e Innovaci\'on (FIS2009-12773-C02-01). E. T. and S. I. acknowledge financial support from the Basque Government (Grants No. BFI08.151 and BFI09.39). X. C. acknowledges support from Juan de la Cierva Programme and the National Natural Science Foundation of China (Grant No. 60806041).
\appendix
\section{Adiabaticity condition for time-dependent non-Hermitian Hamiltonians applied to a decaying two-level atom}
The adiabaticity condition for time-dependent Hermitian Hamiltonians is given by
\beq
\label{adiab_cond}
|\la n(t)|\partial_t m(t) \ra| \ll \frac{1}{\hbar} |E_n(t) - E_m(t)|,\; n\neq m,
\eeq
in terms of instantaneous eigenstates and eigenvalues.
Following closely its derivation in \cite{Schiff}
we generalize it for non-Hermitian Hamiltonians as
\beq
\label{adiab_cond_NH}
|\la \hat{n}(t)| \partial_t{m}(t) \ra| \ll \frac{1}{\hbar} |E_n(t) - E_m(t)|.
\eeq
For the two-level decaying atom this condition is
\beq
|\la \hat{n}_+(t)| \partial_t{n}_-(t) \ra| \ll \frac{1}{\hbar} |E_+(t) - E_-(t)|.
\eeq
Introducing here Eqs. (\ref{E_+_-}), (\ref{n-}) and (\ref{n-1}), the adiabaticity condition for this system takes the form
\beq
2 |\Omega_a(t)| \ll |\Omega(t)|,
\eeq
where $\Omega(t) = \sqrt{-[\Gamma+2i\Delta(t)]^2 + 4\Omega^2_R(t)}$ and $\Omega_a(t)= - \dot{\alpha}/2$.

\end{document}